# In Self-contradiction, Machian Geocentrism Entails Absolute Space

Herbert I. Hartman[1] and Charles Nissim-Sabat[2]

(1) Retired  (2) Northeastern Illinois University, Chicago (*emer.*)

Luka Popov has attempted to advance Machian physics by maintaining that the heliocentric system must be replaced by Tycho Brahe's geocentric system. We show that while geocentrism relies on Mach's contention that accelerations are relative, this contention is untenable because, inter alia, the consequences of an acceleration of an object with respect to the fixed stars cannot be duplicated by acceleration of the stars with respect to this object and, if the universe and a co-rotating observer have the same angular velocity, this motion is detectable because they have different linear velocities. Also, geocentrism precludes the relativity of accelerations and leads to an absolute space with a definite reference point while Mach argued against absolute space, Popov's result that the force exerted by the Earth on the Sun depends on the square of the Sun's mass but is independent of the Earth's mass is paradoxical, and the annual asymmetry in the Cosmic Microwave Background falsifies all geocentric (Ptolemaic or 'Tychonic/Brahean) systems.

## 1. Introduction

Luka Popov has attempted to update Machian physics by arguing for replacing the heliocentric system by a 'Tychonian' a.k.a. 'Brahean' geocentric system where the Sun orbits the Earth but all the other planets orbit the Sun. In his "Newton-Machian analysis of a Neo-tychonian model of planetary motions" (Popov1) Popov introduces a "pseudo force" that maintains the Earth stationary as the Sun orbits the Earth and the planets orbit the Sun. [1] In "The Dynamical Description of the Geocentric Universe" (Popov2) he extends his stationary-Earth model to encompass the apparent diurnal motion of all celestial objects, a necessary extension for Machian theory to be self-consistent.[2]

Mach did not discuss Brahe's work (an unexplained omission) but he maintained that the *Ptolemaic* and the *Copernican* systems are equivalent, stating they are "our interpretation" and thus "equally actual."[3] Mach's basis was his belief that *all* motions, *including accelerations,* are relative. Specifically, he rejected Newton's contention that the centrifugal force exhibited by a liquid in a rotating bucket was due to an interaction with absolute space. Instead, Mach insisted that Newton had the burden of proof to show that the same centrifugal force would not be observed if the bucket were stationary but the masses of the universe were rotating around the bucket and, believing it is self-evident that the same force would be observed, he concluded that all motions are relative.[3]

In 2003 we argued that Mach's arguments and predictions were either paradoxical or falsified by the physics of his time.[4] We showed e.g. that Mach's claim of the equivalence of a rotating bucket in a fixed universe to a fixed bucket in a rotating universe is factually invalid and logically incoherent and that Galileo's observation of a full disk of Venus at specific times when the Sun and Venus were in conjunction falsified the Ptolemaic (but not the Tychonian) system because this observation showed that at these times the Sun was between Venus and the Earth. (We had hesitated to mention this since it is often discussed in elementary texts but we had never seen it discussed in the Machian literature. Yet, well-credentialed Mach followers wrote us that we were wrong about Galileo's falsification of the Ptolemaic system. We convinced them that we were right.) We amplified and extended our arguments to SR [5] and then to GR. [6] We have seen no explanation why in his *Science of Mechanics* [3] Mach mentioned Galileo some 30 times, but he never mentioned any of his astronomical discoveries, not the full disk of Venus, nor the Jovian satellites.

We are glad that Popov has abandoned the Ptolemaic system but this is not enough: we had also pointed out that all observations of stellar aberration and Doppler shifts had falsified any geocentric system long ago. Mach himself acknowledged elsewhere that stellar aberration observations confirmed heliocentric system predictions [7] but he offered no argument that would explain stellar aberration in a geocentric system and neither has anyone else.

Popov acknowledges without elaboration that we have raised arguments against Mach's approach. [1], [2] This lack of elaboration may lead some to believe that our arguments have been proven wrong, either by him or by others. This has not been done. We will show below that Popov's approach exemplifies the difficulties faced by a Machian approach to celestial motions and to physics in general.

## 2. Invalidity of Mach's Theory of the Relativity of Accelerations.

### A. The consequences of acceleration of an object with respect to the fixed stars cannot be duplicated by acceleration of the stars with respect to that object.

Popov relies heavily on Mach's contention that accelerations are relative. Mach claimed that the only unambiguous way one can show that a body is accelerating is in reference to the fixed stars. [8] Yet *accelerometers,* such as a mass held in a chamber by three orthogonal pairs of springs, allow the measurement of the acceleration of a body without referring to the stars, and Mach surely knew about accelerometers and, also, that accelerated charges radiate, [9] but he never mentioned either of these. (Note that, in principle, one can measure the local curvature of spacetime for a body in free fall by replacing each of the six springs by a pair of springs with a mass in the middle and thus obtain the tidal gravitational force along the trajectory of the body.)

Mach could have answered that it is an interaction with the fixed stars that produces the readings in an accelerometer but, in the following, we will demonstrate that the consequences of acceleration of an object with respect to the fixed stars cannot be duplicated by acceleration of the stars with respect to that fixed object.

In discussing the centrifugal force on liquid in a rotating bucket, Mach maintained that one would observe the same centrifugal force on liquid contained in a "fixed" bucket if one made the stars revolve around the bucket, thus predicting that the liquid in the fixed bucket is drawn in a direction opposite to the centripetal acceleration of the stars. Mach did not explain how two liquid elements in the fixed bucket, millimeters away from each other but diametrically equidistant from the axis of revolution of the stars, would undergo exactly equal and opposite accelerations thanks to the revolution of presumably randomly distributed stars astronomical distances away. As shown below, a stationary Earth model faces the same problem. Moreover, Mach would predict the same exact symmetry in the shape of the liquid in each of two buckets an arbitrary distance apart that first rotate simultaneously at arbitrary frequencies *ω* and *ω'* around arbitrarily oriented axes and are then stopped and then the universe is made to rotate (at what frequency?), reproducing the same curvature as before in each of the liquids. What homogenizing process produces the exact uniformity and symmetry in each of the liquids' response and also what motion of the stars produces the different centrifugal forces in the two liquids? Also, can one speak of a 'fixed bucket' or 'a rotating universe' in the context of Machian relativity? [4]

Finally, for Mach, the case of the universe and a co-axial bucket co-rotating with an arbitrary angular velocity is indistinguishable from the case where the two are at rest. Yet the two cases are distinguishable by the magnetic fields produced if the universe and the bucket are electrically charged as well as by the presence of transverse Doppler shifts.

In the same vein as with the rotating bucket, Mach could have maintained that accelerometer behavior (and e.m. radiation also) can be duplicated with a body in either uniform motion or at rest and the fixed stars accelerating. This contention can be disproven:

CASE 1: Consider first a water-containing tub on a horizontal frictionless surface on earth accelerating eastward with acceleration **a** (as measured with Doppler shifts and an accelerometer) with respect to the Earth and to the fixed stars: the water rises to a height $h$ at the west end of the tub (the level of the water constitutes an accelerometer).

Arguing that linear accelerations are relative, Mach would have maintained that if (CASE 2) the tub is initially at rest with respect to the earth but these stars are then accelerated westward with acceleration – **a** with respect to the tub, the water in the tub would presumably rise to the same height $h$ at the west end of the tub [to match the result in CASE 1], but with the water thus moving here in the same direction as the stars' acceleration. [The opposite was obtained in Mach's hypothetical case where the stars revolved around a fixed bucket and the water rose in the direction opposite to the stars' centripetal acceleration.] But, should not the same force that causes the water's westward motion when the stars accelerate make the tub accelerate west as well, so that there is no net acceleration of the tub with respect to the stars?

Also, if the tub is accelerated by the acceleration of the stars, shouldn't the Earth be accelerated as well?

Thus, *the consequences of acceleration of a tub with respect to the fixed stars cannot be duplicated by acceleration of the fixed stars with respect to a tub at rest.* (Using different reasoning, we have shown [4] that one cannot have a fixed bucket in a rotating universe and thus one cannot duplicate with a rotating universe what is observed when the bucket is rotated in a fixed universe.)

Furthermore, concerning again the hypothesis that observations due to acceleration of a body with respect to the fixed stars are duplicated by acceleration of the fixed stars with respect to the body, consider two bodies **A** and **B** initially at rest with respect to the fixed stars and each having an accelerometer showing no acceleration:

**A** remains at rest and its accelerometer shows no acceleration while

**B** suddenly accelerates with respect to **A** and its accelerometer shows acceleration.

No appeal to acceleration of the fixed stars will explain this event. On the other hand, one will find a net force acting only on **B**.

Finally, in Newtonian physics, with objects $S$ and $E$ in an inertial laboratory, one may choose a coordinate system such that the position of $S$ or $E$ is zero. The same is true for the velocity of $S$ or $E$ obtained by the time-derivative of the position. This is Galilean/Newtonian relativity. But we cannot do what relativity of acceleration should allow us to do, i.e. obtain the acceleration of $S$ or $E$ by differentiating again and then choose a coordinate system where the acceleration of either $S$ or $E$ is zero. We cannot do this *because accelerations are produced by forces and measured by accelerometers*. The acceleration of either $S$ or $E$ is zero only when the net force on S or E is zero: Relativity of acceleration violates Newton's Second Law.

**B. The consequences of rotation of an object with respect to the fixed stars cannot be duplicated by revolution of the stars around the fixed object.**

Given the expression for the acceleration observed in a rotating frame of reference:

$$\mathbf{a}_r = \mathbf{a}_i - 2(\boldsymbol{\omega} \times \mathbf{v}_r) - \boldsymbol{\omega} \times (\boldsymbol{\omega} \times \mathbf{r})$$

where *r, i* refer to values in a rotating and an inertial frame of reference, Popov2 formulates a <u>non-relativistic</u> "dynamical" theory of the diurnal motions of the celestial objects[2] by stating that the rotation of the universe with angular velocity $\boldsymbol{\omega}_{rel}$ relative to the origin of the frame, i.e. the Earth, generates a "gravito-magnetic" vector potential **A** at a distance **r** from that origin:

$$\mathbf{A} = \boldsymbol{\omega}_{rel} \times \mathbf{r}$$

Also, Popov maintains that observers on Earth having a relative angular velocity with respect to the universe cannot determine whether the Earth or the universe is rotating, so that his formalism does not entail absolute rotation nor need one consider SR. [2]

A serious shortcoming in the Popov2 reasoning is that his expression for the vector potential supposedly present in a rotating frame within the universe does not depend on the magnitudes or the positions of the masses of the universe, and thus it is not Machian, nor is it "dynamical." The Popov2 formalism fits perfectly within an absolute space theory.

Furthermore, as we have shown, observers on Earth having a relative angular velocity with respect to the universe can determine whether the Earth or the universe is rotating. First, observations with artificial satellites cannot be explained with a stationary non-rotating Earth, most dramatically so in the case of a geosynchronous satellite in an equatorial orbit, which, for a stationary non-rotating Earth, Popov would have it hovering above, say, Singapore, seemingly totally unaffected by gravity, but at only one specific altitude (42,164 km from the Earth's center) while, with a rotating Earth, one observes blue-shifted stellar spectra in the East and red-shifted stellar spectra in the West, both on Earth and on the satellite; [4] Also, SR predicts the observation of stellar transverse Doppler shifts in a rotating universe and, obviously, one has superluminal velocities for most celestial objects. [5] Finally, almost a century ago, it was shown GR predicts a gravito-magnetic centripetal axial force in a rotating universe. [6] Thus gravito-magnetism in a rotating universe is not compatible with Mach's expectations.

**3. Flaws in the Mach/Popov Geocentric system.**

Popov1 first summarizes the Newtonian approach to planetary motion with a single planet (the Earth) orbiting the Sun and notes that

- (a) the Earth/Sun orbital angular momentum is conserved and
- (b) the Earth and the Sun actually orbit around the pair's center of mass. [1]

Popov fails to mention that, with Newton,

- (c) the system's linear momentum is conserved,
- (d) we can consider simultaneously the motion of all the planets using the same laws for all of them,
- (e) for every planet, one can calculate the mass $M$ of the Sun from the angular velocity of the planet's orbit and its mean distance from the Sun (Kepler's Third law), and,
- (f) the ratio of the radii of a planet's and the Sun's orbits around their common center of mass yields the ratio of their masses.

For the Newtonian Earth/Sun system and, with $E, S$ being the Earth and Sun, $r_{SE}$ indicating a vector $r$ directed from S to E, $R$ the S/E distance, $M$ and $m$ the masses of $S$ and $E$, one has for the accelerations of the Earth and the Sun in a heliocentric system:

$$a^H_E = - GM r_{SE}/R^3, \qquad (1)$$

$$a^H_S = Gm\, r_{SE}/R^3. \qquad (2)$$

Mach had stated "the world was given to us only *once* and the Ptolemaic or Copernican view is *our* interpretation, but both are equally actual." [3] The Earth being given to us only once, the same reasoning leads to the conclusion that the flat Earth and the round Earth are equally actual. The flat Earth and the Ptolemaic views have the advantage of presenting simple explanations of *some* sensory data. Neither presents a model that encompasses all the relevant data and the Keplerian/Newtonian system fits the data much better. Given Galileo's observations of Venus, could one have said in 1910, in all seriousness, that the Ptolemaic and the Copernican systems have the same actual validity?

Popov endeavors to put geocentrism on a more formal footing, stating "one is obliged to speak about the *real* forces resulting from the *fact* that the Universe as a whole moves around the observer sitting on the *stationary* Earth." [1] *(Our emphasis).* We have no explanation what these real forces are and why the Universe singles out the Earth to exert on it these forces so as to make it, *alone*, **stationary,** and, **stationary with respect to what?** Does not a stationary Earth imply we have absolute space?

Given an Earth/Sun system with a *stationary* Earth and a gravitational force between S and E, according to Newton ($a=F/m$), the geocentric accelerations $\boldsymbol{a}^G$ are:

$$\boldsymbol{a}^G_E = 0 \quad , \quad (3)$$

$$\boldsymbol{a}^G_S = Gm\boldsymbol{r}_{SE}/R^3. \quad (4)$$

Popov does not mention these because Eq. (4) does not yield the same radius/period relation for S orbiting E as the heliocentric system yields for E orbiting S. Instead he sets out to compute the *"pseudoforce"* which he claims is really responsible for the Sun's acceleration of toward the Earth, enunciating what we call his *'Fundamental Postulate'*:

> *In the Machian picture, the centripetal acceleration is a mere relative quantity, describing the rate of change of relative velocity. Therefore, centripetal acceleration of the Sun with respect to the Earth is given by [the negative of o ur Equation (1)], the acceleration of the Earth in the heliocentric system.*

We have not found any statement by Mach that centripetal accelerations are relative quantities. But he did state: "Two bodies *K'* and *K* that gravitate toward each other impart to each other in the direction of their line of junction accelerations inversely proportional to their masses *m', m*." [10]  Mach did not except the Earth/Sun system from the class of 'two bodies *K'* and *K* that gravitate toward each other.') Also, the *Fundamental Postulate* constitutes a questionable application of the concept of the "relativity of accelerations."  One statement of the latter (Statement X) would be that 'given objects A and B where B accelerates from A with acceleration *b*, one can assign to A and B arbitrary identical collinear accelerations providing that of B is *b* larger than that of A as seen by all observers,' i.e. there is no intrinsic meaning to '*the acceleration of a body*.'  Statement X is contrary to experiment, but it is in the spirit of Galilean relativity. Yet Popov uses a different formulation (Statement Y), where 'A accelerating towards (or away from) B with acceleration *b* is equivalent to B accelerating towards (or away from) A with acceleration –*b,'* a statement that cannot be verified by accelerometers attached to A and B nor by distant observers and that violates Newton's Second and Third Laws.  (Finally, the *Fundamental Postulate* does not take into account that Sun and Earth do not constitute an isolated two-body system and we are not told how this version of the relativity of accelerations can be adapted to the entire solar system.)

Thus the *Fundamental Postulate* does not represent Mach's views and, if it were true, a child twirling a rock at the end of a string would often see herself flying through the air as she orbits the rock.

Furthermore, Eq. (1) pertains to Newton's heliocentric system and we have no explanation why it applies to a geocentric system that violates much of Newton's physics: the use of Eq. (1) discards Newton's Third Law and the Law of Conservation of Momentum. But by the *ad hoc* assignment to the Sun in this geocentric system the same acceleration as that of the Earth in the heliocentric system Popov ensures that the Sun in the geocentric system will have nearly the same orbit as the Earth in the heliocentric system.

Eq. (1) yields for the pseudo-force of the Earth on the Sun, with $R$ being the magnitude of $r_{SE}$ and *with M being the mass of the Sun*:

$$F_{psES} = GM^2 r_{SE}/R^3, \qquad (5)$$

and hence a pseudopotential for an object of mass $m'$ and a position $r'$ with respect to the Earth:

$$U_{ps}(r') = GMm' r' \cdot r_{SE}/R^2. \qquad (6)$$

This reasoning raises further issues. In the geocentric system the centripetal acceleration of the fixed Earth is <u>zero</u> but that of the Sun orbiting the Earth cannot be zero. Relativity of accelerations as used by Popov (Statement Y) allows one to choose either one to be zero and thus relativity of accelerations precludes a geocentric system.

In any event, in principle, we can determine the accelerations of the Sun and the Earth by using observations from space vehicles. Also, the planets orbiting the Sun exert a complex time-varying force on the Earth which must remain stationary. This requires that the Earth be in an infinitely deep potential well.

## 4. Discussion.

Popov gives no explanation how his formalism encompasses stellar aberration and Doppler shifts, the Solar System orbit around the center of our galaxy, the annual variation of the cosmic microwave background radiation, etc… We are not shown how a geocentric system duplicates the heliocentric features (a) through (f) in Sec. 3 that reflect fundamental laws of mechanics. This rejection of what we know is in the spirit of Mach's contention in 1910 that the Ptolemaic and Copernican systems were equally "actual," when, in his *Science of Mechanics,* [3] he cited Galileo's work some thirty times, but never mentioned Galileo's astronomical discoveries. Mach disregarded at least as much evidence against geocentrism as 15th century flat-earth advocates disregarded against the sphericity of the Earth. He acted as a trial lawyer who, addressing a jury, presents all arguments in her client's favor without ever mentioning any evidence pointing in another direction.

Equations (5) and (6) lead to surprising consequences:

These equations do not depend on the mass of the Earth so that the Sun and all other objects would experience the same acceleration if the Earth were but a mere speck of dust, if the Earth were not there at all. We have a singularity fixed in space as the Earth responds to no external gravitational force. We are well on the way to an *absolute space,* where the position of the Earth defines a standard of rest and a reference point from which distances of all other bodies can be measured. This is contrary to Mach's teaching calling absolute space a "monstrous conception." [11] (Singularities in GR respond to external forces.)

In Equation (5) the *centripetal force on the Sun* depends on the *square of the mass of the Sun.* But the Sun is a composite object, with $M=m_1+m_2+...$ (but we must subtract the binding energies between mass components). This is an odd situation, with the force on an object from an external body depending on whether we compute *M* first and then find the total force on the object or we compute the force on each component *m* and then add these individual forces. We know of no other case where the force on an object due to an external body depends on the square of the mass of the object. Eq. (6) tells us that we do not have this problem with any other object. Thus Popov endows the Sun with its own unique mechanics.

The unique mechanics for the Earth and the Sun in this Machian geocentric system depart from what Mach himself characterized as Newton's twofold achievement: the discovery of *universal gravitation* and the formal enunciation of the principles of mechanics. [12]

Finally, Popov invokes Einstein's early interest in Mach to imply that the latter's views are supported by General Relativity. Einstein ended up denying GR supported Mach, denying that GR incorporated the notion that inertia was caused by an interaction between the masses of the universe and specifically denying that relativity of accelerations provided a "workable basis for a new theory."[13]

## 5. Conclusion

The Mach/Popov Tychonian geocentric system is based on a theory of the relativity of accelerations which negates the laws of physics, is easily falsified by experiment, and cannot support geocentrism. Also, in this Tychonian system, the stationary Earth serves as a fixed point and thus entails an absolute space. We are given no inkling what force keeps the Earth stationary, nor any means to reconcile this model with the observations showing the Earth orbits the Sun, the Sun orbits the center of our Galaxy, etc... Finally, the Earth and the Sun are each endowed with its own unique mechanics.*